\documentclass[11pt]{article}

\usepackage{setspace,graphicx,epstopdf,amsmath,amsfonts,amssymb,amsthm,versionPO}
\usepackage{marginnote,datetime,enumitem,subfigure,rotating,fancyvrb}
\usepackage{hyperref,float}
\usepackage{graphicx}
\usepackage{color}
\usepackage{url}
\usepackage{multirow}
\usdate

\usepackage{titlesec}
\newcommand{\periodafter}[1]{#1.}
\titleformat{\subsubsection}[runin]
{\normalfont\bfseries}{\thesubsubsection}{1em}{\periodafter}

\excludeversion{notes}		
\includeversion{links}          

\iflinks{}{\hypersetup{draft=true}}

\ifnotes{%
\usepackage[margin=1in,paperwidth=10in,right=2.5in]{geometry}%
\usepackage[textwidth=1.4in,shadow,colorinlistoftodos]{todonotes}%
}{%
\usepackage[margin=1in]{geometry}%
\usepackage[disable]{todonotes}%
}

\makeatletter\let\chapter\@undefined\makeatother 

\usepackage{endnotes}    
\usepackage{rfs}          

\begin{document}

\setlist{noitemsep}  
\onehalfspacing      

\newcommand{\tit}{Worldcoin: A Hypothetical Cryptocurrency for the People and its Government}

\newcommand{\abs}{The world of cryptocurrency is not transparent enough though it was established for innate transparent tracking of capital flows. The most contributing factor is the violation of securities laws and scam in Initial Coin Offering (ICO) – which is used to raise capital through crowdfunding. There is a lack of proper regularization and appreciation from governments around the world which is a serious problem for the integrity of cryptocurrency market.  We present a hypothetical case study of a new cryptocurrency to establish the transparency and equal right for every citizen to be part of a global system through the collaboration between people and government.  The possible outcome is a model of a regulated and trusted cryptocurrency infrastructure that can be further tailored to different sectors with a different scheme.}

\title{{\bf \tit}}

\author{{\bf Sheikh Rabiul Islam} \\}
\date{\today}

\maketitle

\newpage

\doublespacing
\renewcommand{\footnote}{\endnote}  

\centerline{\Large \bf \tit}
\vspace*{1in}
\centerline{\bf Abstract}
\medskip
\abs

\clearpage
Coin offering industries are growing very quickly due to the blind popularity of cryptocurrency. According to TokenData \cite{5:online}, in the first half of the year 2018, around 309 offerings of \$10.3 billion have been held, compared with 77 offerings of  \$1.2 billion during the same period of the year 2017 \cite{4:online}. Whereas, in the United States, \$15.6 billion raised through the Initial Public Offerings (IPO) in the first three months of 2018 \cite{4:online}. The funding raised by ICO has a tremendous growth rate compared to the IPO counterpart. But some ICOs are scamming investors and exploiting the gray areas of available securities laws with a tokenized version of some sort of product, service, or a promise of future product or investment (e.g., research, infrastructure) \cite{7:online}.

On July 25, 2017, the Security and Exchange Commission (SEC) announced that if the offered token looks like a security then the SEC will recognize and regulate that \cite{7:online}. But Initial Coin Offerings (ICO) equivocates with Initial Public Offerings (IPO) as it is an act of selling money (i.e., coin) instead of security with a promise of future enterprise that will be a worth of a lot of money. And people don’t really worried enough about whether the underlying asset is really linked to something feasible – their assumption is that “someone dumber than them will buy their tokens for more than they paid” \cite{6:online}.  According to Ms. Friedman, the Nasdaq chief executive, coin offerings pose real issues to the average kind of investors \cite{4:online}.  She also expressed her concern about the lack of transparency, oversight, and accountability for those companies that are going out to raise capital through an ICO \cite{6:online}. China has already banned companies from raising money through ICOs and told local regulators to scrutinize 60 of their major ICO platforms \cite{8:online}. Hacking is another threat, in January 2018, hackers broke into Coincheck Inc. (a Japanese cryptocurrency exchange) system and transferred nearly \$500 million in digital tokens (NEM coins) – which is one of the biggest heists in history of digital market, is due to lack of multi-signature security (i.e., requires multiple signatures to transfer funds), though Coincheck attributes this to an insider job \cite{9:online}. Moreover, the bank payment network SWIFT has abandoned the blockchain project concluding that the cost outweighs the benefits that come from it \cite{1:online}. So there is a lot of concerns with cryptocurrency that needs immediate attention. 

Without some kind of appreciation from the government of the participating countries, there is hardly a good future of these crypto markets. When the government will see that the country and the people are really going to be benefitted from this technology in the long run then there might be an appreciation from the government. This interest will also lead them to establish proper regularization for a sustainable crypto market. According to our proposed model, the more and more countries join in the chain the more and more people will be benefitted from a worldwide cryptocurrency market where each citizen will have some participation and each baby born with a fortune. Initially, this fortune might be very negligible, but in times this might be a good asset to fight global hunger, poverty, and inequality. Also, this can be a model of a regulated and trusted cryptocurrency infrastructure that can be tailored to different sectors. And this is the driven force behind our hypothetical case study on the proposed cryptocurrency. 

We named the hypothetical cryptocurrency as Worldcoin. The main purpose of this coin is to ensure each citizen the right of a transparent digital currency which is approved by the government. In the beginning, by default, the government of the participating country owns a number of Worldcoin which is equal to its number of people, each person of the country also owns a Worldcoin. So, in the beginning, the total number of Worldcoin that a country will have for free is exactly twice of its population—half of the total coin is own by the government and remaining half is own by its people. After it’s inauguration, each baby borns with two Worldcoin one goes for its government and one is kept in the name of the newborn which is not redeemable until the baby is at a certain age.  

\begin{figure}[h!]
\centering
\includegraphics[scale=.65]{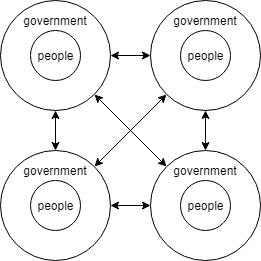}
\caption{Worldcoin transaction channel showing communication paths.}
\label{fig:worldcoin}
\end{figure}

Let’s think about it in a different way then, could this accelerate population growth? Might be, if so then to offset that, penalize the parents after a certain number of the babies from the same parents. Let’s have a look at the bright side of it, a newborn is the owner of a Worldcoin – this might have a positive effect on the importance of the baby in the family and in the government. Do the countries with less population are going to accept it? Well, it is their choice, should they want to get some free cryptocurrency with an active participation in fighting human inequality.  Now what to do with this Worldcoin? 

How will this Worldcoin provide some tangible or intangible benefits? Our idea is that a Worldcoin can be further split into 1000 pieces (each piece is called millicoin) to give it some liquidity and extend its usage to various avenues. Figure \ref{fig:worldcoin}, shows the coin transaction channel of four participating countries.  People can buy or sell Worldcoin or Millicoin on its channel (i.e., with anybody within the country). The government can use its Worldcoins to trade with other governments. Not only trade, any government can send a donation to other government in case of emergency or natural disasters promptly.  There will be a decentralized and distributed ledger or blockchain that will keep track of both the government transactions and people transactions (in some extents). Everything will be transparent to all participating governments. But for local transactions (in people’s channel), the people's privacy can be kept with public-private key encryption. 

Users and governments all over the world can have an idea of how many Worldcoins currently exist and in which portion of the world has how much. Nothing is hidden, everything is transparent here. 
So the total number of Worldcoin NT, in the beginning, can be expressed as follows: 
\begin{equation} \label{eq:formula1}
NT = 2 \times (population \: of \: the \: world)
\end{equation}
Here, our assumption is that all the countries of the world are going to participate in it. But in the real case, this is not guaranteed. So the Formula 1 can be rewritten as follows: 
\begin{equation} \label{eq:formula2}
NT = 2 \times (sum \: of \: populations \: of \: the \: participating \: countries)
\end{equation}

A worldwide balanced economical condition is the driving force to make the population of the country as the base for the number of Worldcoin a country can own freely in the beginning and subsequently with a born of a newborn. Rich are getting richer and poor are getting poorer –  the richest one percent of the American household own 40 percent of the entire country’s wealth and it is higher than it has been at any time in last 50 years \cite{10} \cite{11:online}. 

To commensurate with this outburst, we propose two thresholds in our proposed Worldcoin:  Limit Thresholds (LT) and Cutoff Thresholds (CT). The Cutoff Threshold is a balance (e.g., .1 Worldcoin or 100 millicoin), below which users cannot do any more debit transaction. At this point, the user can only do credit transactions. In other words, a person can never have a balance less than 100 millicoin until a certain age.  The Limit Threshold is a balance (e.g., .5 Worldcoin or 500 millicoin), under which users can sell coins to only those whose balances are not over LT, it can only sell to someone whose balance is also under LT. Though the user can receive from anyone within its channel. We anticipate that forbidding rich people to buy from people who are under LT will squeeze the overall demand but that will be commensurate by the increasing demand (as the price might be little lower due to less number of the buyer) within people under LT. In this way, the overall financial condition of people under LT might improve in this system. This little drop in price might also discourage lower bound people to lose their Worldcoin. 

Similarly, we would like set same two thresholds (with different values) for the government channel too. If a government’s balance is below the Cutoff Threshold then the government can not do any debit transaction. But it can accept credit of Worldcoin from any government. In addition, the Limit Threshold is the balance (e.g., 1 million Worldcoin), below which not allowing the government to sell Worldcoin to governments those are over LT, it can only sell to someone who is also under LT. Though it can receive from any government. As previous, we anticipate that forbidding government to buy from the government who are under LT will squeeze the overall demand but that will be commensurate by the increasing demand (as the price might be little lower due to less number of the buyer) within the governments under LT. In this way, the overall condition of government under LT might improve. We can express these thresholds in the formula as follows:
\begin{equation} \label{eq:formula3}
Balance 	\geqslant ( C1 \times LT) 	\geqslant (C2 \times CT)
\end{equation}
where, C1 and C2 are constant that has different fixed values for the case of people and government channel.

It is worth mentioning that, cryptocurrencies are well known for the slow transaction rate. For example, bitcoin has a transaction rate of 7 per second \cite{3}, which is thousands of times slower compared to some other online transaction medium like VISA which can process 24,000 transactions per second \cite{2:online}. Though our proposed coin will follow a distributed ledger but two types of transactions international (government to government) and local (within peoples of a country) will be handled separately and their transactions are independent of one another. Rather than updating every local transaction in the worldwide ledger, it’s better to update the summary of a group of transaction worldwide in such a way that it is backward traceable by following the individual transaction footprints for forensic purpose. So, we expect to see a good speedup in the transaction. 

It is realistic that not every country is going to start this in this beginning – as this free coins will have no exchange values in the beginning, but in times it will start to have exchange values as it grows. A few countries can start the Worldcoin with the provision of other countries to join at any time in the future. Each country join will have a copy of the distributed ledger. 

Though we represented this hypothetical case study with people and government. Once it sees some success and seems a trusted, regulated and feasible solution then the established cryptocurrency infrastructure and the concept can be tailored to any organizations and within its peoples or stakeholders with different initial offerings. We are far away from a safe and trusted cryptocurrency market, this will not happen in very short time, and without the regularization or appreciation from the government, this might not even see the face of ultimate success. We are not claiming that our proposed solution can solve the problem. But it might show us a direction of the right path towards success. There might be some unknown hurdles, those could be only be perceived by going there. This hypothetical case study might be just the beginning of something good, there might be something bigger branch out from it in times. There are lots of other avenues (e.g., helping disastrous countries) where digital currencies or associated technology can contribute a lot once it is under some sorts of regularization.  There might be something good in it for worldwide humanity and the good thing starts with good people, organization, and government with a good standard of rules, norms, morality, value, and ethics.


\begin{doublespacing}

\bibliographystyle{IEEEtran}
\bibliography{IEEEabrv,bib/references}

\begin{thebibliography}{10}
\providecommand{\url}[1]{#1}
\csname url@samestyle\endcsname
\providecommand{\newblock}{\relax}
\providecommand{\bibinfo}[2]{#2}
\providecommand{\BIBentrySTDinterwordspacing}{\spaceskip=0pt\relax}
\providecommand{\BIBentryALTinterwordstretchfactor}{4}
\providecommand{\BIBentryALTinterwordspacing}{\spaceskip=\fontdimen2\font plus
\BIBentryALTinterwordstretchfactor\fontdimen3\font minus
  \fontdimen4\font\relax}
\providecommand{\BIBforeignlanguage}[2]{{%
\expandafter\ifx\csname l@#1\endcsname\relax
\typeout{** WARNING: IEEEtran.bst: No hyphenation pattern has been}%
\typeout{** loaded for the language `#1'. Using the pattern for}%
\typeout{** the default language instead.}%
\else
\language=\csname l@#1\endcsname
\fi
#2}}
\providecommand{\BIBdecl}{\relax}
\BIBdecl

\bibitem{5:online}
``Token data | news, data and analytics for all ico’s and tokens,''
  \url{https://www.tokendata.io/}, (Accessed on 09/07/2018).

\bibitem{4:online}
``Despite caution over cryptocurrency, investors are bullish - the new york
  times,''
  \url{https://www.nytimes.com/2018/06/27/business/dealbook/cryptocurrency-investors-initial-coin-offerings.html},
  (Accessed on 09/07/2018).

\bibitem{7:online}
``Bitcoin, blockchain, and the trouble with icos | wired,''
  \url{https://www.wired.com/story/ico-cryptocurrency-irresponsibility/},
  (Accessed on 09/07/2018).

\bibitem{6:online}
``The problem with icos is that they’re called icos - mit technology
  review,''
  \url{https://www.technologyreview.com/s/610764/the-problem-with-icos-is-that-theyre-called-icos/},
  (Accessed on 09/07/2018).

\bibitem{8:online}
``Chinese icos: China bans fundraising through initial coin offerings, report
  says,''
  \url{https://www.cnbc.com/2017/09/04/chinese-icos-china-bans-fundraising-through-initial-coin-offerings-report-says.html},
  (Accessed on 09/07/2018).

\bibitem{9:online}
``Coincheck hack: How to steal \$500 million in cryptocurrency | fortune,''
  \url{http://fortune.com/2018/01/31/coincheck-hack-how/}, (Accessed on
  09/07/2018).

\bibitem{1:online}
``Bitcoin and other cryptocurrencies are useless - show me the money,''
  \url{https://www.economist.com/leaders/2018/08/30/bitcoin-and-other-cryptocurrencies-are-useless},
  (Accessed on 09/07/2018).

\bibitem{10}
E.~N. Wolff, ``Household wealth trends in the united states, 1962 to 2016: Has
  middle class wealth recovered?'' National Bureau of Economic Research, Tech.
  Rep., 2017.

\bibitem{11:online}
``The richest 1 percent now owns more of the country’s wealth than at any
  time in the past 50 years - the washington post,''
  \url{https://www.washingtonpost.com/news/wonk/wp/2017/12/06/the-richest-1-percent-now-owns-more-of-the-countrys-wealth-than-at-any-time-in-the-past-50-years/?utm_term=.b11a5a48ebbe},
  (Accessed on 09/07/2018).

\bibitem{3}
S.~Maiyya, V.~Zakhary, D.~Agrawal, and A.~El~Abbadi, ``Database and distributed
  computing fundamentals for scalable, fault-tolerant, and consistent
  maintenance of blockchains,'' \emph{Proceedings of the VLDB Endowment},
  vol.~11, no.~12, 2018.

\bibitem{2:online}
``Small business retail | visa,''
  \url{https://usa.visa.com/run-your-business/small-business-tools/retail.html},
  (Accessed on 09/07/2018).

\end{thebibliography}
\end{doublespacing}

\clearpage
\end{document}